\documentclass[aps,prb,reprint,showpacs]{revtex4-1}
\usepackage[dvipdfmx]{graphicx}
\usepackage{amsmath,amssymb}
\usepackage{bm}

\begin{document}

\title{Microscopic Theory of the Spin Hall Magnetoresistance}

\author{T. Kato$^{1}$, Y. Ohnuma$^{2}$, M. Matsuo$^{2,3,4,5}$}
\affiliation{%
${^1}$Institute for Solid State Physics, The University of Tokyo, Kashiwa, 277-8581, Japan\\
${^2}$Kavli Institute for Theoretical Sciences, University of Chinese Academy of Sciences, Beijing 100190, China\\
${^3}$CAS Center for Excellence in Topological Quantum Computation, University of Chinese Academy of Sciences, Beijing 100190, China\\
${^4}$RIKEN Center for Emergent Matter Science (CEMS), Wako, Saitama 351-0198, Japan\\
${^5}$Advanced Science Research Center, Japan Atomic Energy Agency, Tokai, 319-1195, Japan
}
\date{\today}

\begin{abstract}
We consider a microscopic theory for the spin Hall magnetoresistance (SMR).
We generally formulate a spin conductance at an interface between a normal metal and a magnetic insulator in terms of spin susceptibilities.
We reveal that SMR is composed of static and dynamic parts. The static part, which is almost independent of the temperature, originates from spin flip caused by an interfacial exchange coupling. However, the dynamic part, which is induced by the creation or annihilation of magnons, has an opposite sign from the static part.
By the spin-wave approximation, we predict that the latter results in a nontrivial sign change of the SMR signal at a finite temperature.
In addition, we derive the Onsager relation between spin conductance and thermal spin-current noise.
\end{abstract}
\maketitle 

\section{Introduction}

Magnetoresistance is one of the fundamental phenomena in the research field of spintronics.
Giant magnetoresistance\cite{Baibich1988,Binasch1989,Fert2008} and tunneling magnetoresistance\cite{Julliere1975,Miyazaki1995,Moodera1995,Yuasa2004,Parkin2004} are now essential ingredients in spintronics technology for sensors, memories, and data storage.
Recently, a novel type of magnetoresistance called spin Hall magnetoresistance (SMR) has been observed in a bilayer system composed of a normal metal (NM) and a ferromagnetic insulator (FI)\cite{Nakayama2013a,Chen2013a,Hahn2013,Vlietstra2013,Althammer2013,Meyer2014,Marmion2014,Cho2015,Kim2016,Chen2016,Sterk2019,Tolle2019}.
SMR is explained by the combination of charge-spin conversions in NM and loss of spins at the NM/FI interface\cite{Nakayama2013a,Chen2013a}.
When an in-plane charge current is applied to the NM layer with a large spin-orbit interaction, spin accumulation is caused near the NM/FI interface by the spin Hall effect (SHE).
The amount of spin accumulation is affected by the orientation of FI magnetization because it changes the rate of spin loss at the interface.
A backward spin current owing to spin diffusion is converted into the charge current again by the inverse spin Hall effect (ISHE) and induces longitudinal magnetoresistance, which depends on the orientation of FI magnetization.
The strength of SMR is on the order of $\theta_{\rm SH}^2$, where $\theta_{\rm SH}$ is the spin Hall angle.

Recently, SMR has been reported for the bilayer system composed of NM and an antiferromagnetic insulator (AFI)  \cite{Hou2017,Lin2017,Cheng2019,Hoogeboom2017a,Fischer2018,Ji2018,Lebrun2019}, where the orientation of the N\'eel vector of AFI has been changed by a strong external field or by the orientation of magnetization of FI using the NM/AFI/FI trilayer structure.
The sign of SMR is opposite to the one in the NM/FI bilayer system with respect to the external magnetic field.
This sign change of SMR can be explained if the N\'eel vector of AFI is fixed perpendicular to the external magnetic field or to the magnetization of FI via an exchange bias \cite{Nogues1999,Luan2018}.
We note that a similar sign change of SMR has been observed in a non-collinear-ferrimagnet/NM bilayer system \cite{Ganzhorn2016,Dong2018}.

SMR can be described theoretically by combining the spin diffusion theory with the boundary condition at the interface in terms of the spin-mixing conductance \cite{Nakayama2013a,Chen2013a}. 
However, in this theory, the spin-mixing conductance at the interface is a phenomenological parameter that has to be determined experimentally; therefore, its temperature dependence cannot be predicted.
Furthermore, the magnetization-orientation dependence of the spin-mixing conductance is assumed phenomenologically by its definition.
This semiclassical description of SMR seems to be insufficient for studying quantum features of magnetic insulators such as the effect of thermally excited magnons.
Recently, a microscopic theory of SMR has been proposed based on a local mean-field approach \cite{Zhang2019,Velez2019}. 
However, a general microscopic theory applicable to a wide parameter region is still lacking. 
We note that the physics of SMR is closely related to non-local magnon transport in the NM/FI/NM\cite{Cornelissen2015,Goennenwein2015} and NM/AFI/NM\cite{Lebrun2018} nanostructures.

In addition, the thermal noise of pure spin current at the NM/FI interface has been measured using ISHE \cite{Kamra2014}.
Although the Onsager relation between the thermal noise and spin-mixing conductance at the interface has been discussed qualitatively \cite{Kamra2014}, thus far, it has not been derived by a microscopic theory.
The explicit derivation of the Onsager relation provides an important basis for a theory of nonequilibrium spin-current noise, which generally includes important information on spin transport \cite{Kamra2016a,Kamra2016b,Matsuo2018,Aftergood2018,Joshi2018,Nakata2018,Kato2019} as suggested by studies on electronic current noise \cite{Blanter2000,Thierry2005}.

In this study, we construct a microscopic theory of SMR, which is based on the tunnel Hamiltonian method \cite{Bruus2004,Ohnuma2013}.
We derive a general formula of a spin conductance at the interface for both NM/FI and NM/AFI bilayer systems.
We note that our theory provides the first microscopic description for the NM/AFI bilayers to our knowledge.
By applying the spin-wave approximation, we discuss the temperature dependence of SMR well below the magnetic transition temperature.
In addition, we formulate the spin-current noise in the same framework, and derive the Onsager relation, i.e., the relation between the thermal spin-current noise and spin conductance.

This paper is organized as follows.
In Sec. \ref{sec:SMR}, we theoretically describe SMR by combining the spin diffusion theory in NM and spin conductance at the interface.
In Sec. \ref{sec:model}, we provide the microscopic Hamiltonian for the NM/FI (or NM/AFI) bilayer system.
In Sec. \ref{sec:SpinCurrent}, we formulate the spin conductance at the interface using the tunnel Hamiltonian method. 
In Sec. \ref{sec:SMRFI} and Sec. \ref{sec:SMRAFI}, we discuss the temperature dependence of spin conductance using the spin-wave approximation.
In Sec. \ref{sec:experiment}, 
we briefly discuss the relevance of this study to the SMR experiments.
In Sec.\ref{sec:noise}, we formulate the spin current noise and explicitly derive the Onsager relation.
Finally, we summarize our results in Sec. \ref{sec:summary}.
The details of the derivation are provided in two appendices.

\section{Theoretical Description of SMR}
\label{sec:SMR}

\begin{figure}[!tb]
\begin{center}
\includegraphics[width=80mm]{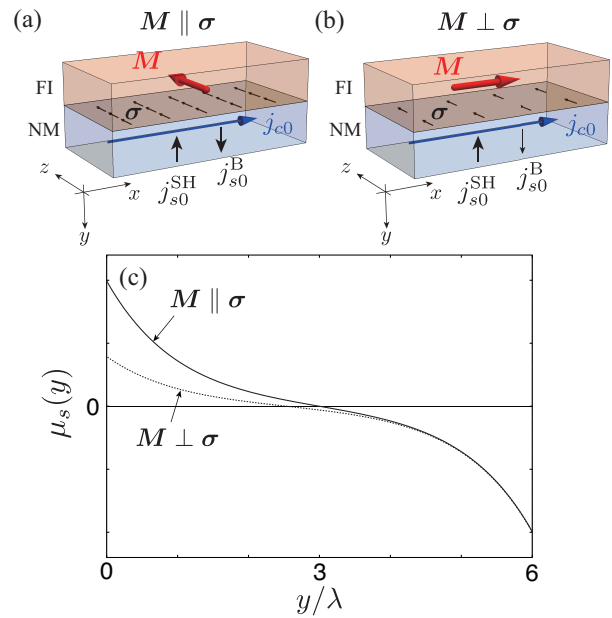}
\caption{(Color online) Schematic diagram of the normal-metal(NM)/ferromagnetic-insulator(FI) bilayer system for the spin Hall magnetoresistance (SMR) measurement.
When an external charge current is applied to NM in the $x$-direction, a spin current $I_s$ with a $z$ component flows in the $y$-direction owing to the spin Hall effect.
Spin current induces spin accumulation ${\bm \sigma}$ at the NM/FI interface with magnetization ${\bm M}$.
(a) Parallel configuration of ${\bm \sigma} \parallel {\bm M}$.
(b) Perpendicular configuration of ${\bm \sigma} \perp {\bm M}$.
(c) Chemical potential difference $\delta \mu_s = \mu_{\uparrow} - \mu_{\downarrow}$, which is defined for a quasi-equilibrium steady state, is shown as a function of position $y/\lambda$, where
$\lambda$ is the spin diffusion length, and the thickness of NM is set as $d_N = 6 \lambda$.
The NM/FI interface is located at $y=0$.
}
\label{fig_setup}
\end{center}
\end{figure}

We theoretically describe the SMR by improving the spin diffusion theory provided in Ref.~\onlinecite{Chen2013a}.
First, let us consider a bilayer system composed of NM and FI layers.
When we apply electric field to the NM layer in the $+x$ direction, a spin current $j_{s0}^{\rm SH}=\theta_{\rm SH} \sigma E_x$ is driven in the $-y$ direction by the spin Hall effect, where $\theta_{\rm SH}$ is the spin Hall angle, $\sigma$ is the electric conductivity, and $E_x$ is the $x$ component of electric field.
Here, we have defined the spin current $j_{s0}^{\rm SH}$ as a difference between charge currents of opposite spins.
This spin current induces spin accumulation at the interface between NM and FI layers, as shown in Fig.~\ref{fig_setup}(a) and (b).
In the abovementioned figure, the direction of spins accumulated at the interface is denoted with ${\bm \sigma}$.
In a steady state, the spin current $j_{s0}^{\rm SH}$ is balanced with a backflow spin current 
\begin{align}
j_s^{\rm B} = -(\sigma/2e)\partial_y \mu_s(y) .
\end{align}
Here, $\mu_s(y)$ is the spin chemical potential defined as
\begin{align}
\mu_s(y) = \mu_\uparrow(y)-\mu_\downarrow(y) .
\end{align}
Using the continuity equation and accounting for spin relaxation, we show that $\mu_s(y)$ obeys the following differential equation 
\begin{align}
\frac{d^2\mu_s}{dy^2} = \frac{\mu_s}{\lambda^2} ,
\label{eq:diffeq}
\end{align}
where $\lambda$ is the spin diffusion length.
The spin chemical potential $\mu_s(y)$ is obtained as a function of $y$ by solving this equation under the boundary conditions
\begin{align}
j_s^{\rm B}(y) - j_{s0}^{\rm SH} = \left\{ \begin{array}{ll} -j_s^{\rm I}, & (y=0), \\
0, & (y=d_N), \end{array} \right.
\label{eq:diffmus}
\end{align}
where $d_N$ is the thickness of the NM layer, and $j_s^{\rm I}$ is the spin absorption rate at the NM/FI interface that depends on the direction of magnetization of FI.

In this study, we use another definition of spin current at the interface. 
We define $I_s$ as a decay rate of spin angular momenta at the NM/FI interface. 
This can be related with $j_s^{\rm I}$ as follows:
\begin{align}
j_s^{\rm I} = \frac{e}{\hbar/2} \frac{I_s}{S}, 
\end{align}
where $S$ is the cross section area of the NM/FI interface.
We define a dimensionless spin conductance as
\begin{align}
G_s = \lim_{\mu_s(0)\rightarrow 0} \frac{I_s}{\mu_s(0)},
\label{eq:defGs}
\end{align}
where $\mu_s(0)$ is the chemical potential difference at the NM/FI interface.

The solution of the differential equation~(\ref{eq:diffeq}) is written in the form of $\mu_s(y)=Ae^{-y/\lambda}+Be^{y/\lambda}$.
We note that the constants, $A$ and $B$, include $\mu_s(0)$ through the boundary condition, Eq.~(\ref{eq:diffmus}), by approximating the spin current as $I_s \simeq G_s \mu_s(0)$.
By solving the self-consistent equation for $\mu_s(0)$, we obtain
\begin{align}
\mu_s(0) = \frac{\mu_{s0}}{1+g_s \coth(d_N/\lambda)}
\end{align}
where $\mu_{s0}$ is the chemical potential difference in the absence of the NM/FI interface, and $g_s$ is the dimensionless factor, which is defined as
\begin{align}
g_s = \frac{4e^2}{\hbar} \frac{G_s}{\sigma S/\lambda},
\end{align}
and indicates the amplitude of the absorption rate at the interface.

In Ref.~\onlinecite{Chen2013a}, the magnetization-orientation dependence of $G_s$ was discussed in terms of the spin mixing conductance.
In this discussion, the spin absorption rate at the interface, $g_s$, is largest (smallest) when the magnetization ${\bm M}$ is perpendicular (parallel) to the accumulated spin ${\bm \sigma}$ (see Fig.~\ref{fig_setup}(a) and (b)).
Then, the spatial profile of $\mu_s(y)$ changes depending on the direction of ${\bm M}$ (see Fig.~\ref{fig_setup}(c)).
A similar approach was employed also in recent theoretical works on unidirectional SMR\cite{Sterk2019} and low-dimensional-FI/NM systems\cite{Velez2019}.
In our study, however, no assumption is made about the magnetization-orientation dependence of $G_s$.
As shown later, the magnetization-orientation dependence of the spin absorption rate, which is implicitly assumed in the discussion that is based on the mixing conductance, is not sufficient to discuss the temperature dependence of the SMR signal.

The backflow current $j_s^{\rm B}$ induces a charge current in the $x$ direction owing to the inverse spin Hall effect.
Then, longitudinal magnetoresistance is calculated as\cite{Chen2013a}
\begin{align}
\frac{\Delta \rho}{\rho} = \theta_{\rm SH}^2 \frac{g_s  \tanh^2(d_N/2\lambda)}{1+g_s \coth(d_N/\lambda)} .
\label{SMRformula}
\end{align}
Thus, SMR is related to the spin conductance $G_s$ via the factor $g_s$.
In the subsequent sections, we calculate $G_s$ as a function of the angle between ${\bm M}$ and ${\bm \sigma}$.

The theoretical description of SMR for the NM/AFI bilayer is almost the same as for the NM/FI bilayer.
SMR can be discussed by calculating $G_s$ as a function of an angle between the N\'eel vector and ${\bm \sigma}$.
We note that the present formulation is applicable to more complex systems such as a NM with the Rashba spin-orbit interaction\cite{Tolle2019}.

\section{Model}
\label{sec:model}

In this section, we introduce a model for NM/FI and NM/AFI bilayers.
After we provide the Hamiltonian for bulk systems of NM (Sec.~\ref{sec:modelNM}), FI (Sec.~\ref{sec:modelFI}), and AFI (Sec.~\ref{sec:modelAFI}), we describe the model of the interfacial exchange coupling in Sec.~\ref{sec:modelINT}.

\subsection{Normal Metal}
\label{sec:modelNM}

The Hamiltonian for a bulk NM is given as
\begin{align}
H_{\rm NM} &= \sum_{{\bm k}\sigma} \xi_{\bm k} c^\dagger_{{\bm k}\sigma} c_{{\bm k}\sigma}, 
\label{eq:HamiltonianNM}
\end{align}
where $\xi_{\bm k}=\epsilon_{\bm k}-\mu$ is the energy dispersion measured from the chemical potential, and $\sigma = \uparrow,\downarrow$ is the $z$ component of an electron spin.
We assume that the spin accumulation at an interface induced by SHE is described by quasi-thermal equilibrium states with spin-dependent chemical potential shifts, $\pm \mu_s/2$, where $\mu_s$ is recognized as its value at the interface, $\mu_s(0)$, given in the previous section.
The density matrix for this quasi-thermal equilibrium state is given as $\rho=e^{-\beta {\cal H}_{\rm NM}}/Z$, where
\begin{align}
{\cal H}_{\rm NM} &= 
\sum_{{\bm k}\sigma} (\xi_{\bm k}-\sigma \mu_s/2) c^\dagger_{{\bm k}\sigma}  
c_{{\bm k}\sigma} ,
\label{eq:HamiltonianNM2}
\end{align} 
where $\beta$ is the inverse temperature, and $Z={\rm Tr}(e^{-\beta {\cal H}})$ is the ground partition function.

\subsection{Ferromagnetic Insulator (FI)}
\label{sec:modelFI}

\begin{figure}[!tb]
\begin{center}
\includegraphics[width=75mm]{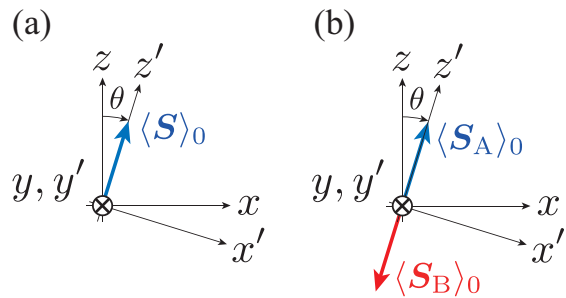}
\caption{Relation between the magnetization-fixed coordinate $(x,y,z)$ and laboratory coordinate $(x',y',z')$ for (a) FI and (b) antiferromagnetic insulator (AFI).}
\label{fig:rotation}
\end{center}
\end{figure}

For the Hamiltonian of bulk FI, we consider the Heisenberg model given as
\begin{align}
H_{\rm FI} &= J \sum_{\langle i,j \rangle}
{\bm S}_i \cdot {\bm S}_j - \hbar \gamma h_{\rm dc} \sum_{i} S^{z'}_{i},
\label{HFI} 
\end{align}
where ${\bm S}_j$ is the localized spin, $J$ ($<0$) is the ferromagnetic exchange coupling, 
$\langle i,j \rangle$ indicates a pair of nearest-neighbor sites,
$\gamma$ is the gyromagnetic ratio, and $h_{\rm dc}$ is the static magnetic field.
Here, we introduce a new coordinate $(x',y',z')$ and assume that the net magnetization is aligned in the $+z'$ direction in this new coordinate by the magnetic field (see Fig.~\ref{fig:rotation}(a)):
\begin{align}
\langle {\bm S}_j \rangle_0
= (\langle S^{x'}_j \rangle_0,
\langle S^{y'}_j \rangle_0,
\langle S^{z'}_j \rangle_0)
= (0,0,\tilde{S}_0),
\label{eq:magnetization}
\end{align}
where $\langle \cdots \rangle_0$ indicates the thermal average in bulk FI, and $\tilde{S}_0$ is the amplitude of the magnetization per site, which depends on the temperature.

\subsection{Antiferromagnetic Insulator}
\label{sec:modelAFI}

For the Hamiltonian of bulk AFI, we consider the Heisenberg model on a lattice composed of two sublattices, A and B:
\begin{align}
H_{\rm AFI} &= J \sum_{\langle i, j \rangle} {\bm S}_{{\rm A},i} \cdot {\bm S}_{{\rm B},j},
\end{align}
where ${\bm S}_{\nu,j}$ denotes a localized spin on the sublattice $\nu$ ($={\rm A},{\rm B}$); the antiferromagnetic exchange is denoted as $J$ ($>0$), and $\langle i,j \rangle$ indicates the nearest-neighbor pairs between two sublattices.
We assume that the magnetization on the sublattice A(B) is aligned in the $+z'$($-z'$) direction (see Fig.~\ref{fig:rotation}(b)):
\begin{align}
& \langle {\bm S}_{{\rm A},j} \rangle
= (\langle S_{{\rm A},j}^{x'} \rangle, \langle S_{{\rm A},j}^{y'} \rangle, \langle S_{{\rm A},j}^{z'} \rangle)
= (0,0,\tilde{S}_0), 
\label{eq:msa} \\
& \langle {\bm S}_{{\rm B},j} \rangle
= (\langle S_{{\rm B},j}^{x'} \rangle, \langle S_{{\rm B},j}^{y'} \rangle, \langle S_{{\rm B},j}^{z'} \rangle)
= (0,0,-\tilde{S}_0),
\label{eq:msb}
\end{align}
where $\tilde{S}_0$ is the amplitude of staggered magnetization per site, which depends on the temperature.

\subsection{Exchange coupling at the interface}
\label{sec:modelINT}

The interfacial exchange coupling between FI (or AFI) and NM is described by the Hamiltonian
\begin{align}
H_{\rm ex} &= \sum_{\nu} \sum_{\langle i,j \rangle} \left[ T_{ij}^{\nu} S_{\nu,i}^+ s_j^- + (T_{ij}^\nu)^* S_{\nu,i}^- s_j^+ \right] ,
\label{eq:HamINT}
\end{align}
where 
$S_{\nu,i}^\pm= S^x_{\nu,i} \pm S^y_{\nu,i}$ are creation and annihilation operators of the spin in the laboratory coordinate, $s_j^\pm$ are those of the spin of conduction electrons, $T_{ij}^{\nu}$ is an exchange coupling between a pair of interfacial sites, $\langle i,j \rangle$, and $\nu$ is the sublattice of localized spins.
The sublattice is unique ($\nu = {\rm A}$) for FI, and there are two sublattices ($\nu = {\rm A}, {\rm B}$) for AFI.

To proceed with the calculation, we need to rewrite the Hamiltonian (\ref{eq:HamINT}) in the spin operators in the magnetization-fixed coordinate $(x',y',z')$.
The conversion formula for the spin operators from the magnetization-fixed coordinate $(x',y',z')$ to the laboratory coordinate $(x,y,z)$ is given as
\begin{align}
S^{x}_{\nu,j} &= \cos \theta \, S_{\nu,j}^{x'} - \sin \theta \, S_{\nu,j}^{z'} , \\
S^{y}_{\nu,j} &= S_{\nu,j}^{y'}, \\
S^{z}_{\nu,j} &= \sin \theta \, S_{\nu,j}^{x'} + \cos \theta \, S_{\nu,j}^{z'} ,
\end{align}
where $\theta$ is the angle of magnetization (see Fig.~\ref{fig:rotation}).
By this coordinate transformation, we obtain 
\begin{align}
S^+_{\nu,j} &= \cos^2 (\theta/2) S^{+\prime}_{\nu,j} -\sin^2 (\theta/2)S^{-\prime}_{\nu,j} - \sin \theta S^{z'}_{\nu,j}, \label{eq:calSjplus} \\
S^-_{\nu,j} &= \cos^2 (\theta/2)S^{-\prime}_{\nu,j} -\sin^2 (\theta/2)S^{+\prime}_{\nu,j} - \sin \theta S^{z'}_{\nu,j},
\label{eq:calSjminus}
\end{align}
where $S^{\pm\prime}_{\nu,j} = S^{x'}_{\nu,j}\pm S^{y'}_{\nu,j}$.
Then, the interface exchange interaction can be divided into three parts as
\begin{align}
H_{\rm ex} &= \sum_{a=1}^3 H_{\rm ex}^{(a)}, \\
H_{\rm ex}^{(a)} &= g_a(\theta) \sum_{\langle i,j \rangle}
\left[ T^\nu_{ij} S_{\nu,i}^{(a)} s_j^-
+  (T^\nu_{ij})^* (S_{\nu,i}^{(a)})^\dagger s_j^+ \right], 
\end{align}
where $S^{(a)}$ and $g_a(\theta)$ ($a=1,2,3$) are defined as
\begin{align}
& S^{(1)}_{i,\nu}=S^{z'}_{\nu,i}, \quad g_1(\theta) = -\sin \theta, 
\label{S1} \\
& S^{(2)}_{i,\nu}=S^{+\prime}_{\nu,i} , \quad g_2(\theta) = \cos^2 (\theta/2), 
\label{S2} \\
& S^{(3)}_{i,\nu}=S^{-\prime}_{\nu,i} , \quad g_3(\theta) =-\sin^2 (\theta/2).
\label{S3} 
\end{align}

\section{Spin Current}
\label{sec:SpinCurrent}

Next, we calculate the spin current using the second order perturbation with respect to the exchange coupling at the interface \cite{Bruus2004,Ohnuma2013,Matsuo2018,Kato2019,Ominato2019,Ominato2020}.
We derive a general formula for spin current and spin conductance.
Our formula is expressed in terms of spin susceptibilities in NM and FI (or AFI) layers and is general, i.e., the formula does not depend on a specific model.

\subsection{Definition}

We define the spin current as the absorption rate of the $z$ component of the spin angular momenta in the NM side at the interface:
\begin{align}
\hat{I}_s &= -\hbar \partial_t s^{z}_{\rm tot} = i [s_{\rm tot}^{z},H], \\
s^{z}_{\rm tot} &= \frac{1}{2}
\sum_{\bm k}(c_{{\bm k}\uparrow}^\dagger c_{{\bm k}\uparrow} - c_{{\bm k}\downarrow}^\dagger
c_{{\bm k}\downarrow}),
\end{align}
where $H=H_{\rm NM} + H_{{\rm FI/AFI}} + H_{\rm ex}$ is the total Hamiltonian.
The spin current is calculated in the form
\begin{align}
\hat{I}_s &= \sum_{a=1}^3 \hat{I}_S^{(a)}, \\
\hat{I}_s^{(a)} &
= -i g_a(\theta) \sum_{\langle i,j\rangle} \sum_{\nu} (T^\nu_{ij} S_{\nu,i}^{(a)} s_j^- - {\rm h.c.}).
\end{align}
We note that the $z$ component of the total spin is not conserved at the interface because the magnetization of FI (or AFI) is not necessarily aligned in the $z$ direction.

\subsection{Second-order perturbation}

We calculate the spin current within the second-order perturbation with respect to the interfacial exchange coupling.
For simplicity, we assume that the correlation between the exchange coupling for different pairs vanishes after random averaging on the positions of the interfacial sites:
\begin{align}
\langle T_{ij}^\nu (T_{i'j'}^{\nu'})^* \rangle_{\rm av} = |T_{ij}^\nu|^2 
\delta_{i,i'} \delta_{j,j'} \delta_{\nu,\nu'},
\label{eq:localapproximation}
\end{align}
Then, the spin current is written as a sum of independent spin exchange processes at different pairs of the interfacial sites.
We note that this kind of assumption has been used for long time to describe electric tunnel junctions\cite{Bruus2004,Kato2019,footnoteHT}.
Then, the spin current is calculated as~\cite{Bruus2004,Kato2019}
\begin{align}
     & \langle \hat{I}_s^{(a)} \rangle 
    =  \hbar^2 g_a(\theta)^2 \sum_{\nu} A_\nu \int_{-\infty}^{\infty} \frac{d\omega}{2\pi} \, 
    \left(-\frac{1}{{\cal N}{\cal N}_{\rm F}} \right)\sum_{{\bm k},{\bm q}} 
    \nonumber \\
    &  \times {\rm Re} \, \Biggl[ \chi^<({\bm q},\omega) G_{\nu \nu}^{R,(a)}({\bm k},\omega)
    + \chi^A({\bm q},\omega) G_{\nu \nu}^{<,(a)}({\bm k},\omega)\Biggr],
\end{align}
where $A_\nu = 2\sum_{\langle i,j\rangle}|T_{ij}^\nu|^2/\hbar$, ${\cal N}$ is the number of sites in NM, and ${\cal N}_{\rm F}$ is the number of unit cells in FI (or AFI).
Hereafter, we set $A= A_{\rm A}$ for FI and assume that the exchange coupling is equivalent for the two sublattices, $A=A_{\rm A}=A_{\rm B}$, for AFI.
The advanced and lesser spin susceptibilities of NM, $\chi^A({\bm q},\omega)$ and $\chi^<({\bm q},\omega)$, are defined by the Fourier transformation of the following functions:
\begin{align}
\chi^A({\bm q},t) &= \frac{i}{{\cal N}\hbar}
    \theta(-t) \langle [ s^+_{\bm q}(t), s^-_{\bm q}(0) ] \rangle_0, \\
    \chi^{<}({\bm q},t) &= \frac{i}{{\cal N}\hbar}
    \langle s^-_{\bm q}(0) s^+_{\bm q}(t) \rangle_0, 
\end{align}
where $\langle \cdots \rangle_0$ indicates an average for the unperturbed Hamiltonian,
$s^{\pm}_{\bm q}$ is the spin operator of conduction electrons given as
\begin{align}
s_{\bm q}^- & = \sum_{\bm k} c_{{\bm k}\downarrow}^\dagger c_{{\bm k}+{\bm q}\uparrow}, \\
s_{\bm q}^+ & = \sum_{\bm k} c_{{\bm k}+{\bm q}\uparrow}^\dagger c_{{\bm k}\downarrow}, 
\end{align}
and $s^{\pm}_{\bm q}(t)=e^{iH_{\rm NM}t/\hbar} s^{\pm}_{\bm q} e^{-iH_{\rm NM}t/\hbar}$.
For quasi-thermal equilibrium states, we can prove the dissipation-fluctuation theorem  
\begin{align}
   \chi^<({\bm q},\omega) &=
   -2i f(\hbar \omega + \mu_s) {\rm Im} \, \chi^A({\bm q},\omega), 
   \label{eq:FDrelation}
\end{align}
using the Lehman representation, where $f(E)=(e^{\beta E}-1)^{-1}$ is the Bose distribution function.
The retarded and lesser spin correlation functions of FI (or the AFI), $G_{\nu \nu'}^{R,(a)}({\bm k},\omega)$, and $G_{\nu\nu'}^{<,(a)}({\bm k},\omega)$, are defined by the Fourier transformation of the following functions:
\begin{align}
& G_{\nu\nu'}^{R,(a)}({\bm k},t) 
 = -\frac{i}{\hbar} \theta(t)\langle 
    [S^{(a)}_{\nu,{\bm k}}(t),(S^{(a)}_{\nu'{\bm k}}(0))^\dagger]
    \rangle_0, \label{GRa} \\
& G_{\nu\nu'}^{<,(a)}({\bm k},t) = -\frac{i}{\hbar} 
\langle (S^{(a)}_{\nu',{\bm k}}(0))^\dagger S^{(a)}_{\nu,{\bm k}}(t) \rangle_0, 
\label{GLa}
\end{align}
where $S^{(a)}_{\nu,{\bm k}}$ ($a=1,2,3$) are the spin operators defined by the spatial Fourier transformation of Eqs.~(\ref{S1})-(\ref{S3}).
To continue calculation, we introduce a fluctuating part of the spin correlation function as $\delta S_{\nu,{\bm k}}^{(a)}(t) =  S_{\nu,{\bm k}}^{(a)}(t) - \langle S_{\nu,{\bm k}}^{(a)} \rangle_0$.
From Eqs.~(\ref{GRa}) and (\ref{GLa}), we obtain \begin{align}
G^{R,(a)}_{\nu \nu}({\bm k},\omega) &= \delta G^{R,(a)}_{\nu \nu}({\bm k},\omega), 
\label{deltaGr}\\
G^{<,(a)}_{\nu \nu}({\bm k},\omega) &= -\frac{2\pi i {\cal N}_{\rm F} \tilde{S}_0^2}{\hbar} \delta_{a,1} \delta_{{\bm k},{\bm 0}} \delta(\omega) \nonumber \\
& + \delta G^{<,(a)}_{\nu \nu}({\bm k},\omega).
\end{align}
where $\delta G^{R,(a)}_{\nu \nu}({\bm k},\omega)$ and $\delta G^{<,(a)}_{\nu \nu}({\bm k},\omega)$ are the correlation functions, which are defined by replacing $S_{\nu,{\bm k}}^{(a)}(t)$ with $\delta S_{\nu,{\bm k}}^{(a)}(t)$ in Eqs.~(\ref{GRa}) and (\ref{GLa}).
Using the Lehman representation, we can prove the dissipation-fluctuation theorem  
\begin{align}
\delta G_{\nu\nu'}^{<,(a)}({\bm k},\omega) &= 2 i f(\hbar \omega) {\rm Im} \, \delta G^{R,(a)}_{\nu\nu'}({\bm k},\omega).
\label{eq:GFD}
\end{align}
Combining Eqs.~(\ref{eq:FDrelation}), (\ref{deltaGr})-(\ref{eq:GFD}), the spin current is calculated as
\begin{align}
\langle \hat{I}_s \rangle &= I_{s,1} + I_{s,2},
\label{eq:formula1} 
\\
I_{s,1} &= \hbar A\sin^2 \theta  \,  \tilde{S}_0^2 N_{\nu} \, {\rm Im} \, \chi_{\rm loc}^R(0), 
\label{eq:formula0} 
\\
I_{s,2}  &= \sum_{a=1}^3 2g_a(\theta)^2 \hbar A \sum_{\nu} \int \frac{d(\hbar \omega)}{2\pi} \, {\rm Im} \, \chi_{\rm loc}^R(\omega) \nonumber \\
& \times (-{\rm Im}\, \delta G_{\nu\nu,{\rm loc}}^{R,(a)}(\omega)) [ f(\hbar \omega)-f(\hbar \omega + \mu_s)],
\label{eq:formulaa} 
\end{align}
where $N_\nu$ is the number of sublattices ($N_\nu = 1$ for FI and $2$ for AFI).
The local spin correlation functions, $\chi_{\rm loc}^{R}(\omega)$ and $G_{\nu\nu,{\rm loc}}^{R,(a)}(\omega)$, are defined as 
\begin{align}
\chi_{\rm loc}^{R}(\omega) &= \frac{1}{\cal N}
\sum_{\bm q} \chi^{R}({\bm q},\omega), \\
G_{\nu\nu,{\rm loc}}^{R,(a)}(\omega) &= \frac{1}{{\cal N}_{\rm F}}
\sum_{\bm k} G_{\nu\nu}^{R,(a)}({\bm k},\omega).
\end{align}

Finally, let us summarize the physical meaning of the spin current formula given by Eqs.~(\ref{eq:formula1})-(\ref{eq:formulaa}).
We stress that this formula is written in terms of spin susceptibilities for bulk materials and, therefore, is applicable to general systems such as NM with strong electron-electron  interactions.
The spin current is composed of two parts.
The first part, $I_{s,1}$, describes the static part, which is induced by spin flipping owing to the static effective transverse magnetic field via interfacial exchange coupling.
Actually, the static part $I_{s,1}$ is almost independent of the temperature well below the magnetic transition temperature and reaches the maximum when the accumulated spin at the interface in the side of NM is perpendicular to the magnetization of FI (or the N\'eel vector of AFI), i.e., $\theta = \pi/2$.
This feature of $I_{s,1}$ coincides with the theory that is based on the spin mixing conductance in Refs.~\onlinecite{Nakayama2013a,Chen2013a}.
However, there exists an additional contribution $I_{s,2}$, which is induced by the creation or annihilation of magnons.
This part can be regarded as a dynamic part.
In subsequent sections, we will show that this dynamic part depends on the temperature and that its angle dependence differs from the static part.

\subsection{Spin conductance}

From Eqs.~(\ref{eq:formula1})-(\ref{eq:formulaa}), the (dimensionless) spin conductance at the interface defined in Eq.~(\ref{eq:defGs}) is calculated as
\begin{align}
G_{s} &= G_{s,1} + G_{s,2}, \label{eq:Gsformula1} \\ 
G_{s,1} & = G_0 \sin^2 \theta, 
\label{eq:Gsformula2} \\
G_{s,2} & = \sum_{a=2}^3 2G_0 g_a(\theta)^2 \frac{1}{N_{\nu}} \sum_\nu \int \frac{dE}{2\pi} \, E \nonumber \\
& \times \left(-\frac{1}{\tilde{S}_0^2} {\rm Im}\, \delta G_{\rm loc}^{R,(a)}(E/\hbar)\right) \left[- \frac{df}{dE}\right],
\label{eq:Gsformula3} 
\end{align}
where $G_0 = \hbar A \tilde{S}_0^2 N_{\nu} \pi N(0)^2$, and $N(0)$ is the density of states in NM at the Fermi energy.
We note that the spin chemical potential $\mu_s(0)$ is now recognized as $\mu_s$ in the model of NM (see Eq.~(\ref{eq:HamiltonianNM2})).

\section{SMR in FI/NM bilayers}
\label{sec:SMRFI}

In this section, we calculate the spin conductance by employing the spin-wave approximation within description of noninteracting magnons for the ferromagnetic Heisenberg model.
Hereafter, we assume that the amplitude of spins for the ground state, $S_0=\tilde{S}(T=0)$, is sufficiently large, and that the temperature is much lower than the magnetic transition temperature.

By applying the Holstein-Primakoff transformation to the spin operators in the magnetization-fixed coordinate, the Hamiltonian of FI is approximately written as
\begin{align}
H_{\rm FI} &= \sum_{{\bm k}} \hbar \omega_{\bm k} b_{\bm k}^\dagger b_{\bm k}, \\
\hbar \omega_{\bm k} &\simeq {\cal D}k^2 + E_0,
\end{align}
where $b_{\bm k}$ ($b_{\bm k}^\dagger$) is the magnon annihilation (creation) operator,
$E_0 = \hbar \gamma h_{\rm dc}$ is the Zeeman energy, and ${\cal D}=|J|S_0a^2$.
In the spin-wave approximation neglecting 
magnon-magnon interaction, the local spin susceptibility is calculated as ${\rm Im} \, \delta G^{R,(1)}_{\rm loc}(E/\hbar) = 0$, and
\begin{align}
& {\rm Im} \, \delta G^{R,(2)}_{\rm loc}(E/\hbar) =
-{\rm Im} \, \delta G^{R,(3)}_{\rm loc}(-E/\hbar)  \nonumber \\
& \hspace{5mm} = -2\pi S_0 D_{\rm F}(E), 
\end{align}
where $D_{\rm F}(E)$ is the density of states for magnon excitation per unit cell:
\begin{align}
D_{\rm F}(E) = \frac{1}{{\cal N}_{\rm F}} \sum_{\bm k} \delta(E - \hbar \omega_{\bm k}).
\end{align}
Although the magnetization $\tilde{S}_0$ depends weakly on the temperature within the present spin-wave approximation, we neglect it for simplicity $(\tilde{S}_0\simeq S_0)$.
Then, the spin conductance takes the form
\begin{align}
G_s &= G_{s,0} + \Delta G_s  \sin^2 \theta , 
\label{GSform}
\end{align}
where $G_{s,0}$ is the part that is independent of $\theta$, and the second term describes the angle dependence, i.e., SMR.
The amplitude of SMR is calculated as
\begin{align}
\Delta G_s &= G_0 (1-g_{\rm F}(T)), 
\label{GSform2} \\
g_{\rm F}(T) & = \frac{1}{S_0} \int_0^{\infty} dE\, E D_{\rm F}(E) \left[- \frac{df}{dE}\right] .
\end{align}
We note that the first term $G_0$ in Eq.~(\ref{GSform2}) originates from the static part $G_{s,1}$, while the second term $-G_0 g_{\rm F}(T)$ originates from the dynamic part $G_{s,2}$.
The factor $g_{\rm F}(T)$ is small under the condition $S_0 \gg 1$, for which the spin-wave approximation based on non-interacting magnon picture works well.
If we neglect $g_{\rm F}(T)$, we recover the usual positive SMR behavior, $\Delta G_s =G_0 \sin^2 \theta$.
The factor $g_{\rm F}(T)$ weakens positive SMR.
When the Zeeman energy is neglected ($E_0 \simeq 0$), the temperature dependence of the SMR signal is obtained at sufficiently low temperatures as 
\begin{align}
\frac{\Delta G_s}{\Delta G_s(T=0)} \simeq 1-\frac{5.2}{S_0} \left( \frac{k_{\rm B}T}{E_{\rm c}} \right)^{3/2} ,
\end{align}
where $E_{\rm c} = {\cal D}k_{\rm c}^2$ is the cutoff energy, which is on the order of the transition temperature (for details, see Appendix~\ref{app:cutoff}).
If $g_{\rm F}(T)$ exceeds 1, the sign of SMR changes.
The temperature of the sign change, $T_{\rm r}$, is estimated as 
\begin{align}
k_{\rm B}T_{\rm r} \sim \left(\frac{S_0}{5.2}\right)^{2/3} E_{\rm c}.
\label{eq:TrFI}
\end{align}
We note that for $S_0 \gg 1$, $T_{\rm r}$ becomes on order of $E_{\rm c}$ for which the non-interacting magnon approximation is not justified.
This estimate indicates that the sign change of SMR occurs if $S_0$ is not large.

\section{SMR in AFI/NM bilayers}
\label{sec:SMRAFI}

In the spin-wave approximation, the Hamiltonian for AFI is obtained in the leading order of $1/S_0$ as
\begin{align}
H_{\rm AFI} &= \sum_{{\bm k}} \hbar \omega_{\bm k} (\alpha_{\bm k}^\dagger \alpha_{\bm k}
+ \beta_{\bm k}^\dagger \beta_{\bm k}), 
\label{AFISW}
\end{align}
where $\alpha_{\bm k}$ and $\beta_{\bm k}$ are the annihilation operators for magnons, $\omega_{\bm k} = v_{\rm m} |{\bm k}|$ is the dispersion relation, and $v_{\rm m} = zJS_0a/(\sqrt{3} \hbar)$ is the velocity of magnons (see 
Appendix~\ref{app:SWAAFI}). 
Here, we approximated the magnon dispersion as a liner dispersion in the long wavelength limit ($|{\bm k}|\rightarrow 0$).
The local spin susceptibility is calculated as ${\rm Im} \, \delta G_{\nu\nu,{\rm loc}}^{R,(1)}(E/\hbar) = 0$ and
\begin{align}
& \sum_{\nu={\rm A},{\rm B}} {\rm Im} \, \delta G_{\nu\nu,{\rm loc}}^{R,(2)}(E/\hbar)= -\sum_{\nu={\rm A},{\rm B}} {\rm Im} \, \delta G_{\nu\nu,{\rm loc}}^{R,(3)}(-E/\hbar) \nonumber \\
& \hspace{5mm} = -2\pi S_0 F(E) (D_{\rm AF}(E)-D_{\rm AF}(-E)),  \label{GAFloc}
\end{align}
where $D_{\rm AF}(E)$ and $F(E)$ are the density of states for magnon excitation and form factor, respectively (see Appendix~\ref{app:SWAAFI}):
\begin{align}
D_{\rm AF}(E) &= \frac{1}{{\cal N}_{\rm F}} \sum_{\bm k}  \delta(E - \hbar \omega_{\bm k}), \\
F(E) &= \frac{\sqrt{3}\hbar v_{\rm m}}{|E|a}.
\label{eq:DAFdef}
\end{align}
Using these results, the spin conductance is calculated in the form of Eq.~(\ref{GSform}).
Then, the amplitude of the SMR is given as
\begin{align}
\Delta G_s &= G_0 (1-g_{\rm AF}(T)) , \\
g_{\rm AF}(T) & =  \frac{1}{S_0} \int_0^{\infty} dE\, E F(E) D_{\rm AF}(E) \left[- \frac{df}{dE}\right] .
\end{align}
For AFI, the temperature dependence of the SMR signal is given as
\begin{align}
\frac{\Delta G_s}{\Delta G_s (T=0)} & \simeq 1-\frac{4.4}{S_0} \left( \frac{k_{\rm B}T}{E_{\rm c}} \right)^{2}, 
\label{eq:SMRtempAFI}
\end{align}
where $E_{\rm c} = \hbar v_{\rm m} k_{\rm c}$ is the cutoff energy (see Appendix~\ref{app:cutoff}).
The temperature of the sign change is estimated as 
\begin{align}
k_{\rm B}T_{\rm r} \sim \left(\frac{S_0}{4.4}\right)^{1/2} E_{\rm c}.
\label{eq:TrAFI}
\end{align}
This estimate indicates that the sign change of SMR may occur if $S_0$ is not large.

\section{Comparison with Experiments}
\label{sec:experiment}

Let us first consider SMR for FI.
SMR experiments for FI have been performed mainly for Pt/YIG bilayer systems~\cite{Nakayama2013a,Chen2013a,Vlietstra2013,Althammer2013,Hahn2013}.
In this theory, the temperature of the sign change of SMR signal estimated for Pt/YIG exceeds the magnetic transition temperature using Eq.~(\ref{eq:TrFI}) and $S_0=10$.
This indicates that the correction by the factor of $g_{\rm F}(T)$ is small, and no sign change occurs.
This result is consistent with the detailed measurement of of SMR in Pt/YIG\cite{Meyer2014,Marmion2014}, where the observed temperature dependence is interpreted by that of the spin diffusion length.
Our result also provides insights into measurement of non-local magnetoresistance in Pt/YIG/Pt nanostructures\cite{Cornelissen2015,Goennenwein2015}, which is induced by magnon diffusion in YIG; our result indicates that the temperature dependence of non-local magnetoresistance mainly comes from that of the spin diffusion in YIG.

\begin{figure}[!tb]
\begin{center}
\includegraphics[width=75mm]{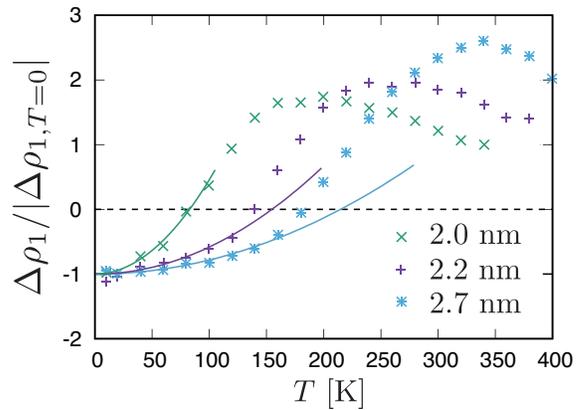}
\caption{(Color online) Experimental data of SMR (amplitude of the magnetization-dependent part of magnetoresistance) measured in Pt/NiO/YIG bilayer systems for the NiO thickness of 2.0, 2.2, and 2.7 nm; obtained from Ref.~\onlinecite{Hou2017}.
The solid curves show fitting using quadratic temperature dependence described by Eq.~(\ref{eq:SMRtempAFI}).
The data is normalized as $-1$ at zero temperature using the extrapolated value by the fitting.
}
\label{fig:compare}
\end{center}
\end{figure}

SMR was also measured for AFI/NM bilayer systems in several experiments.
In Fig.~\ref{fig:compare}, we show the measured temperature dependence of SMR in Pt/NiO/YIG bilayer systems; obtained from Ref.~\onlinecite{Hou2017}. 
In this experiment, an estimate of the factor $g_s \coth(d_N/\lambda)$ is much smaller than 1 using $\lambda=1.5 \ {\rm nm}$, $d=4 \ {\rm nm}$, , $\sigma^{-1} = 860 \ \Omega \cdot {\rm nm}$, and $G_s/(\hbar/2e^2)\sim 3\times 10^{12} \ \Omega^{-1} \cdot {\rm m}^{-2}$.
Therefore, SMR is proportional to spin conductance (see Eq.~(\ref{SMRformula})).
As seen from Fig.~\ref{fig:compare}, the sign of SMR changes at $80$, $140$, and $180 \ {\rm K}$ for NiO thickness of 2.0, 2.2, and 2.7 nm.
In addition, we show the fitted curve using quadratic temperature dependence described by Eq.~(\ref{eq:SMRtempAFI}) in the figure.
This fitting indicates that the quadratic temperature dependence  explains well the experimental data at low temperatures.
If we employ $S_0=0.94$ and $E_{\rm c} = 1500 \ {\rm K}$ from the magnon dispersion measured by the neutron experiment~\cite{Hutchings1972}, the temperature of this sign change is estimated for bulk NiO as $T_{\rm r} = 690 \ {\rm K}$ (see also Appendix~\ref{app:cutoff}).
This estimated temperature for the sign change is much larger than the experimental observation shown in Fig.~\ref{fig:compare}.
However, the N{\'e}el temperature of NiO ($T_N = 525 \ {\rm K}$) is suppressed for the thin layer~\cite{Alders1998}, which indicates a decrease in the magnon velocity.
In addition, the non-interacting magnon approximation holds well only at low temperatures compared to the N{\'e}el temperature.

In this paper, we discussed SMR at low temperatures using the spin-wave approximation neglecting magnon-magnon interaction.
Because the spin current formula derived in this paper is general, SMR can be evaluated for arbitrary temperatures using a numerical method such as the Monte Carlo method.
Detailed numerical analysis beyond the non-interacting magnon approximation and the consideration of roughness of the interface is left as a future problem.

\section{Onsager Relation}
\label{sec:noise}

In this section, we formulate noise in thermal equilibrium ($\delta \mu_s = 0$) and derive the Onsager relation, which relates thermal noise to spin conductance.

We define the noise power of the spin current as
\begin{align}
{\cal S} =\int_{-\infty}^{\infty} dt (\langle \hat{I}_{\rm S}(t) \hat{I}_{\rm S}(0) \rangle+\langle \hat{I}_{\rm S}(0) \hat{I}_{\rm S}(t) \rangle).
\end{align}
In second-order perturbation with respect to the exchange coupling at the interface, we can replace the average with that for an unperturbed system as $\langle \cdots \rangle \simeq \langle \cdots \rangle_0$.
Using a similar procedure that is used for spin current, the noise power is calculated as
\begin{align}
{\cal S} & = {\cal S}_1 + {\cal S}_2, \\
{\cal S}_1 &= 2\hbar^2  A \tilde{S}_0^2 N_\nu \sin^2 \theta \nonumber \\
& \hspace{5mm} \times \lim_{\omega \rightarrow 0} (-i) [\chi_{\rm loc}^<(\omega) + \chi_{\rm loc}^>(\omega) ], \\
{\cal S}_2 &= \sum_{a=1}^3 2\hbar^2 A g_a(\theta)^2 \sum_{\nu} \int_{-\infty}^{\infty}\frac{d(\hbar \omega)}{2\pi} \nonumber \\
& \times \Biggl[\chi_{\rm loc}^<(\omega) \delta G_{\nu\nu,{\rm loc}}^{>,(a)}(\omega) + \chi^>(\omega) \delta G_{\nu\nu,{\rm loc}}^{<,(a)}(\omega) \Biggr].
\end{align}
Here, $\chi_{\rm loc}^>(\omega)$ and $G_{\nu\nu',{\rm loc}}^{>,(a)}(\omega)$ are the greater components of local spin susceptibilities defined as
\begin{align}
& \chi^>_{\rm loc}(\omega) =\frac{i}{{\cal N}^2\hbar} \sum_{\bm q} \int dt \, e^{i\omega t} \langle  s_{\bm q}^+(t) s_{\bm q}^-(0) \rangle ,  \\
& \delta G^{>,(a)}_{\nu \nu',{\rm loc}}(\omega) = -\frac{i}{{\cal N}_{\rm F}\hbar} \sum_{\bm k} \int dt \, e^{i\omega t} \nonumber \\
& \hspace{25mm} \times \langle \delta S_{\nu,{\bm k}}^{(a)}(t) (\delta S_{\nu',{\bm k}}^{(a)}(0))^\dagger \rangle_0, 
\end{align}
Using the dissipation-fluctuation relations
\begin{align}
\chi_{\rm loc}^>(\omega) &= 2 i (1+f(\hbar \omega+\delta \mu_s)) {\rm Im} \, \chi_{\rm loc}^R(\omega),  \\
\delta G_{\nu\nu',{\rm loc}}^{>,(a)}(\omega) &= 2 i (1+f(\hbar \omega)) {\rm Im} \, \delta G_{\nu\nu',{\rm loc}}^{R,(a)}(\omega),
\end{align}
the thermal spin-current noise is calculated as
\begin{align}
{\cal S}_1 &= 
 2 \hbar k_{\rm B} T G_0 \sin^2 \theta,  \\
{\cal S}_2 &= \sum_{a=1}^3 8 \hbar^2 A g_a(\theta)^2 \sum_{\nu} \int_{-\infty}^{\infty}\frac{d\omega}{2\pi} \,
{\rm Im} \, \chi^R({\bm q},\omega)  \nonumber \\
& \hspace{5mm} \times {\rm Im} \, G_{\nu \nu,{\rm loc}}^{R,(a)}({\bm k},\omega) 
\,  f(\hbar \omega)(1+ f(\hbar \omega) ).
\end{align}
Using the identity
\begin{align}
f(\hbar \omega)(1+ f(\hbar \omega)) = k_{\rm B}T \left(-\frac{df}{dE}\right),
\end{align}
and comparing these results with Eqs.~(\ref{eq:Gsformula1})-(\ref{eq:Gsformula3}), we can prove the Onsager relation
\begin{align}
{\cal S} = 4 \hbar k_{\rm B} T G_s.
\end{align}
We stress that this proof is general, and this relation holds at arbitrary temperatures for a wide class of the Hamiltonian for NM and FI (or AFI).

\section{Summary}
\label{sec:summary}

We constructed a microscopic theory for spin Hall magnetoresistance observed in bilayer systems composed of a normal metal and a ferromagnetic (or antiferromagnetic) insulator.
We formulated the spin current at the interface in terms of spin susceptibilities and clarified that it is composed of static and dynamic parts.
The static part of spin current originates from spin flip owing to an effective magnetic field induced by an interfacial exchange coupling.
This part is almost independent of the temperature, and takes a maximum when the magnetization (or the N{\'e}el vector) is perpendicular to accumulated spins in a normal metal, which is consistent with intuitive discussions in previous experimental studies.
However, the dynamic part, which is induced by creation or annihilation of magnons, depends on the temperature, and has opposite magnetization dependence, i.e., takes a maximum when the magnetization (or the N{\'e}el vector) is parallel to accumulated spins in a normal metal.
The dynamic part becomes larger when the amplitude of the localized spin, $S_0$, is smaller.
This indicates that the sign of SMR changes at a specific temperature if $S_0$ is sufficiently small.
Our study gives the first microscopic description of SMR in the NM/AFI bilayer systems.
We also discussed that the measured temperature dependence of the SMR in the Pt/NiO/YIG trilayer system~\cite{Hou2017} is consistent with our results.
Finally, we proved the Onsager relation between spin conductance and thermal spin-current noise using a microscopic calculation.

Our general formalism, which is applicable to various systems for arbitrary temperatures, is essential for describing spin Hall magnetoresistance.
Theoretical analysis beyond the non-interacting magnon approximation and the extension of our thoery toward non-collinear magnets are left as future problems.

\acknowledgements
T.K. is financially supported by JSPS KAKENHI Grant Numbers JP20K03831.
M.M. is financially supported by the Priority Program of Chinese Academy of Sciences, Grant No. XDB28000000 and KAKENHI (No. 20H01863) from MEXT, Japan.

\appendix

\section{Spin-Wave Approximation for AFI}
\label{app:SWAAFI}

In this appendix, we briefly summarize the spin-wave approximation neglecting the magnon-magnon interaction for AFI based on non-interacting magnons.
Throughout this appendix, the amplitude of spins for the ground state, $S_0=\tilde{S}(T=0)$, is much larger than 1, and the temperature is much lower than the magnetic transition temperature.

By standard procedure based on the Holstein-Primakoff transformation, the Hamiltonian of AFI is approximately written as
\begin{align}
H_{\rm AFI} &\simeq J z S_0 \sum_{{\bm k}} \left( a_{{\bm k}}^\dagger \ b_{{\bm k}} \right)
\left( \begin{array}{cc} 1 & \zeta_{\bm k} \\ \zeta_{\bm k}^* & 1 \end{array} \right)
\left( \begin{array}{c} a_{{\bm k}} \\ b_{{\bm k}}^\dagger \end{array} \right), 
\end{align}
where $a_{\bm k}$ and $b_{\bm k}$ are the annihilation operators of spins on A and B sublattices, respectively, and $z$ is the number of nearest neighbor sites.
For the cubic lattice ($z = 6$), $\eta_{\bm k}$ is calculated as
\begin{align}
\zeta_{\bm k} &= \frac13 (\cos k_x a + \cos k_y a + \cos k_z a ),
\end{align}
To diagonalize the Hamiltonian, we introduce the Bogoliubov transformation:
\begin{align}
& a_{\bm k} = u_{\bm k} \alpha_{\bm k} - v_{\bm k} \beta_{\bm k}^\dagger, \\
& b_{\bm k}^\dagger = v_{\bm k} \alpha_{\bm k} - u_{\bm k} \beta_{\bm k}^\dagger.
\end{align}
The exchange relation of bosonic operators leads to the constraint $u_{\bm k}^2-v_{\bm k}^2=1$.
By straightforward calculation, we obtain the diagonalized Hamiltonian (\ref{AFISW}) in the main text using the solutions
\begin{align}
u_{\bm k}^2 = v_{\bm k}^2+1 = \frac{1}{2} \left(\frac{1}{\sqrt{1-\zeta_{\bm k}^2}}+1 \right),
\end{align}
In the long-wavelength limit (${\bm k} \rightarrow 0$), the dispersion relation becomes
$\omega_{\bm k} = v_{\rm m} |{\bm k}|$, where 
$v_{\rm m} = JzS_0 a/(\sqrt{3}\hbar)$ is the velocity of magnons.
Using this diagonalized Hamiltonian, the local spin susceptibilities are written as Eq.~(\ref{GAFloc}) with the density of state of magnons given by Eq.~(\ref{eq:DAFdef}).
Spin susceptibilities include the factor $v_{\bm k}^2 + u_{\bm k}^2$, which is rewritten in the limit of ${\bm k} \rightarrow 0$ by the form factor
\begin{align}
F(E) = \frac{1}{\sqrt{1-\zeta(E)^2}} = \frac{JzS_0}{E},
\end{align}
where $\zeta(E) = \zeta_{\bm k}|_{\hbar \omega_{\bm k} = E}$.

\section{Cutoff Energy}
\label{app:cutoff}

To estimate the temperature at which the sign of SMR changes, we approximate magnon dispersion as that for the continuum limit (${\bm k} \rightarrow 0$).
We introduce the cutoff wavenumber, $k_{\rm c}$, as ${\cal N}_{\rm F} = V(2\pi)^{-3} \int_0^{k_c} dk \, 4\pi k^2$, leading to $k_{\rm c} a =(6\pi^2)^{1/3}\equiv \alpha$,
where $a$ is the lattice constant.
The density of states of magnons is calculated for FI using its cumulative function defined as
$D^{\rm c}_{\rm F}(E) = ((E-E_0)/E_{\rm c})^{3/2}$,
where $E_{\rm c}={\cal D}k_{\rm c}^2$.
When we neglect the Zeeman energy $E_0$, we obtain 
\begin{align}
D_{\rm F}(E) &= \frac{dD^{\rm c}_{\rm F}(E)}{dE} = \frac{3}{2E_{\rm c}}
\left(\frac{E}{E_{\rm c}}\right)^{1/2}.
\end{align}
For AFI, the cumulative function is given as
$D^{\rm c}_{\rm AF}(E) =(E/E_{\rm c})^3$,
where $E_{\rm c} = \hbar v_{\rm m} k_{\rm c}$. 
We 
obtain the density of state of magnons as
\begin{align}
D_{\rm AF}(E) = \frac{dD_{\rm AF}^{\rm c}}{dE} = \frac{3}{E_{\rm c}} \left(\frac{E}{E_{\rm c}}\right)^2.
\end{align}

To estimate SMR for the Pt/NiO/YIG bilayer system, we used the parameters obtained from the neutron scattering experiment~\cite{Hutchings1972}.
The antiferromagnetic structure of NiO is described by four sublattices, each of which is a simple cubic lattice with the strongest antiferromagnetic coupling, $J\simeq230 \ {\rm K}$, for nearest neighboring sites, which leads to $E_{\rm c}=1500 \ {\rm K}$.

\bibliography{mybibfile}

\end{document}